\documentclass[10pt, conference, letterpaper]{IEEEtran}
\IEEEoverridecommandlockouts
\usepackage{cite}
\usepackage{amsmath,amssymb,amsfonts}
\usepackage{algorithmic}
\usepackage{graphicx}
\usepackage{textcomp}
\usepackage[dvipsnames,svgnames]{xcolor}
\def\BibTeX{{\rm B\kern-.05em{\sc i\kern-.025em b}\kern-.08em
    T\kern-.1667em\lower.7ex\hbox{E}\kern-.125emX}}

\usepackage{balance}

\usepackage{multirow}
\usepackage{booktabs}
\usepackage{pgfplots}

\usepackage{caption}
\usepackage[position=b]{subcaption}

\usepackage{dblfloatfix} 
\graphicspath{{./figs/}}

\usepackage{url}

\usepackage{tikz}
\usetikzlibrary{calc,arrows,arrows.meta,external,colorbrewer,matrix,shapes,shapes.misc}

\usepackage[absolute,showboxes]{textpos}

\newcommand{\timepoint}{circle (1.5pt)}

\usepackage{pifont}

\usepackage{xspace}

\newcommand{\eg}{\textit{e.g.,}~}
\newcommand{\ie}{\textit{i.e.,}~}

\newcommand*\circled[1]{\tikz[baseline=(char.base)]{
            \node[shape=circle,draw,inner sep=0.9pt] (char) {\sf \small #1};}}

\makeatletter
\renewcommand{\paragraph}[1]{\vspace*{0.03in}\noindent{\bf #1.}\hspace{0.25ex \@plus1ex \@minus.2ex}}
\makeatother

\tikzexternaldisable

\IEEEoverridecommandlockouts

\IEEEpubid{\makebox[\columnwidth]{978-3-903176-47-8~\copyright 2022 IFIP \hfill} \hspace{\columnsep}\makebox[\columnwidth]{ }}

\begin{document}
\bstctlcite{IEEEexample:BSTcontrolNew}

\setlength{\TPHorizModule}{\paperwidth}
\setlength{\TPVertModule}{\paperheight}
\TPMargin{5pt}
\begin{textblock}{0.8}(0.1,0.02)
	\noindent
	\footnotesize
	\centering
	If you cite this paper, please use the TMA reference:
	R. Hiesgen, M. Nawrocki, T. C. Schmidt, and M. Wählisch.
	2022. The Race to the Vulnerable: Measuring the Log4j Shell Incident.
	\emph{In Proc. of Network Traffic Measurement and Analysis Conference (TMA ’22).}
	IFIP, 9 pages. 
\end{textblock}

\title{The Race to the Vulnerable: \\ Measuring the Log4j Shell Incident}

\author{
  \IEEEauthorblockN{Raphael Hiesgen}
  \IEEEauthorblockA{HAW Hamburg\\raphael.hiesgen@haw-hamburg.de}
 \and
  \IEEEauthorblockN{Marcin Nawrocki}
  \IEEEauthorblockA{Freie Universit\"at Berlin\\marcin.nawrocki@fu-berlin.de}
 \and
  \IEEEauthorblockN{Thomas C. Schmidt}
  \IEEEauthorblockA{HAW Hamburg\\t.schmidt@haw-hamburg.de}
 \and
  \IEEEauthorblockN{Matthias W\"ahlisch}
  \IEEEauthorblockA{Freie Universit\"at Berlin\\m.waehlisch@fu-berlin.de}
 }

\maketitle

\begin{abstract}
The critical remote-code-execution (RCE) Log4Shell is a severe vulnerability that was disclosed to the public on December 10,~2021. It exploits a bug in the wide-spread Log4j library. Any service that uses the library and exposes an interface to the Internet is potentially vulnerable.

In this paper, we measure the rush of scanners during the two months after the disclosure. We use several vantage points to observe both researchers and attackers. For this purpose, we collect and analyze payloads sent by benign and malicious communication parties, their origins, and churn. We find that the initial rush of scanners quickly ebbed. Especially non-malicious scanners were only interested in the days after the disclosure. In contrast, malicious scanners continue targeting the vulnerability.
\end{abstract}


\section{Introduction}

The world runs on software. With that comes a constant threat of new vulnerabilities which affect widely deployed implementations. In the past, Heartbleed (2014) \cite{dlkab-mh-14,CVE-2014-0160}, Rowhammer~(2015) \cite{CVE-2015-0565}, and Spectre (2018) \cite{khfgg-saese-20,CVE-2017-5753,CVE-2017-5715} disrupted the operation of systems around the world. The most recent vulnerability in this line is Log4Shell (2021). It enables remote code execution through vulnerable applications by injecting a prepared string into the omnipresent Log4j~library.

For hours, days, or weeks after the disclosure of such a vulnerability a race starts between multiple parties. Attackers want to exploit systems before they get patched. Operators and other companies need to update their systems before they get compromised. Security firms monitor malicious activities and researchers aim to understand details and asses the scale of the vulnerability.  
Releasing a patch and protective measures alongside a public disclosure is a necessary step to enable admins to secure their systems, but no guarantee exists for a timely implementation.

As an integral part of Java applications, Log4jShell can be exploited in a variety of application-dependent ways. We focus our observations on network requests. We collect and analyze data over a two months period using reactive network telescopes at four vantage points on two continents. 

After recapitulating the events close to the Log4Shell disclosure in Section~\ref{sec:bg} and introducing our data set in Section~\ref{sec:method}, we trace the scanners in Section~\ref{sec:scanners}. We find that scanning started on the day of the disclosure and spiked about a week later with a focus on HTTP related ports. Next, in Section~\ref{sec:payloads} we examine the collected payloads before focusing on the URLs central to the exploit in Section~\ref{sec:urls}.  %
Finally, Section~\ref{sec:conclusion} discusses and concludes our findings.

\section{Background and Related Work}
\label{sec:bg}

The Log4Shell vulnerability (CVE-2021-44228~\cite{CVE-2021-44228}) in the popular Log4j library was publicly disclosed on Dec~10,~2021 by Apache alongside a fix in Log4j library version 2.15.0. The bug allows for remote code execution~(RCE) by injecting prepared strings into the logging library. NIST~\cite{nist-nvd-44228} classified this threat as a critical vulnerability with the highest severity rating.
An impact assessment by Google~\cite{g-uialv-21} estimates that 4\% (or 17,000 packages) on the Maven Central were affected, either directly or via dependencies. This is twice the impact of an average packet (mean 2\%, median 0.1\%), which highlights the popularity of Log4j.

From a high level perspective, the exploit works by injecting a formatted message into the logging component, which then interprets and executes the message. Specifically, the Log4j library supports format messages that are evaluated by the library and can be used to add additional information to log messages, such as the Java version. One of the services that can be used for runtime evaluations is JNDI~\cite{a-jlps-13}, the Java Naming and Directory Interface, which in turn can query a variety of lookup services. Log4Shell focuses on the services LDAP and RMI for injecting code and infecting the local machine.

\subsection{A Brief History of the Log4Shell Incident}
\label{sub:timeline}

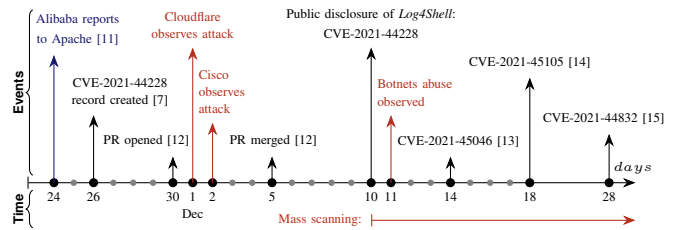
\begin{figure}
\begin{tikzpicture}[
	>=Stealth,
	font=\tiny,
	good/.style={MidnightBlue},
	bad/.style={BrickRed},
	thiswork/.style={BurntOrange}
	]
\draw[|->] (-10pt,0) coordinate (start) -- ++(230pt,0) node[above,xshift=-1pt] {$days$};
\foreach \x in {0,7.5,15,...,180}{
	\draw[fill,gray] (\x pt,0) circle (1pt);
}

\foreach \x in {210}{
	\draw[fill,gray] (\x pt,0) circle (1pt);
}

\draw[fill] (0pt,0) node[below] {24} \timepoint;
\draw[->,good] (0pt,0) -- ++(0,1.7) node[above,xshift=9pt] {\shortstack{Alibaba reports\\to Apache~\cite{a-alsv-22}}};

\draw[fill] (15pt,0) node[below] {26} \timepoint;
\draw[->,] (15pt,0) -- ++(0,.9) node[above,xshift=10pt] {\shortstack{CVE-2021-44228\\record created~\cite{CVE-2021-44228}}};

\draw[fill] (45pt,0) node[below] {30} \timepoint;
\draw[->,] (45pt,0) -- ++(0,0.35) node[above,xshift=-10pt] {\shortstack{PR opened~\cite{g-rlavj-21}}};

\draw[fill] (52.5pt,0) node[below] {\shortstack{1\\Dec}} \timepoint;
\draw[->,bad] (52.5pt,0) -- ++(0,1.8) node[above] {\shortstack{Cloudflare\\observes attack}};

\draw[fill] (60pt,0) node[below] {2} \timepoint;
\draw[->,bad] (60pt,0) -- ++(0,0.8) node[above,xshift=3.5pt]{\shortstack[l]{Cisco\\observes\\attack}};

\draw[fill] (82.5pt,0) node[below] {5} \timepoint;
\draw[->,] (82.5pt,0) -- ++(0,0.35) node[above,xshift=.5pt] {\shortstack{PR merged~\cite{g-rlavj-21}}};

\draw[fill] (120pt,0) node[below] {10} \timepoint;
\draw[->,] (120pt,0) -- ++(0,1.8) node[above] {\shortstack{Public disclosure of \textit{Log4Shell}:\\CVE-2021-44228}};

\draw[fill] (127.5pt,0) node[below] {11} \timepoint;
\draw[->,bad] (127.5pt,0) -- ++(0,0.9) node[above,xshift=9pt] {\shortstack[l]{Botnets abuse\\observed}};

\draw[fill] (150pt,0) node[below] {14} \timepoint;
\draw[->,] (150pt,0) -- ++(0,.35) node[above,xshift=3pt] {\shortstack{CVE-2021-45046~\cite{CVE-2021-45046}}};

\draw[fill] (180pt,0) node[below] {18} \timepoint;
\draw[->,] (180pt,0) -- ++(0,1.4) node[above] {\shortstack{CVE-2021-45105~\cite{CVE-2021-45105}}};
s
\draw[fill] (210pt,0) node[below] {28} \timepoint;
\draw[->,] (210pt,0) -- ++(0,.65) node[above,xshift=-2pt] {\shortstack{CVE-2021-44832~\cite{CVE-2021-44832}}};

\draw[|->,bad] (120pt, -0.5) node[left] (top rel) {Mass scanning:} -- ++(100pt,0);

;
\draw[decorate,decoration=brace] (-8pt,-0.6) -- node[sloped,above] {\sffamily\bfseries{Time}} ++(0,0.5);

\draw[decorate,decoration=brace] (-8pt,.1) -- node[sloped,above] {\sffamily\bfseries{Events}} ++(0,2.2);
\end{tikzpicture}
\tikzexternalenable
\caption{The unfolding of the Log4Shell vulnerability from reporting to the consequences in 2021.}
\label{fig:timeline}
\end{figure}

Figure~\ref{fig:timeline} shows the timeline of observations around the Log4Shell incident. The vulnerability was originally reported to Apache on Nov~24 by the Alibaba Cloud Security Team~\cite{a-alsv-22}. The Chinese-based firm quickly faced consequences of reporting the vulnerability directly to Apache~\cite{scmp-albci-21,z-crsac-21} instead of contacting national authorities first. A CVE record was created on Nov~26~\cite{CVE-2021-44228} but not published until public disclosure on Dec~10. In the meantime, a pull request~(PR) to address the vulnerability was opened on Nov~30~\cite{g-rlavj-21} and merged five days later. Cloudflare notes that they observed the first exploit one day after the PR on Dec~1st~\cite{p-cf-tweet}. Cisco observed an exploitation attempt a day later on Dec~2nd, as published in their Talos Blog~\cite{ct-tacal-21}. Both companies report that widespread scanning started on Dec~10.

Within a day of the public release, the Mirai and Muhstik botnets, crypto miners, and other malware were observed to use the exploit for propagation~\cite{nl-talvh-21,j-lapw-21}. Microsoft further reports that nation-state attackers experiment with the exploit and integrate it into their activities~\cite{m-gpdhe-21}.

The first fix was insufficient and more exploits followed (CVE-2021-45046~\cite{CVE-2021-45046}, CVE-2021-45105~\cite{CVE-2021-45105}, CVE-2021-44832~\cite{CVE-2021-44832})---none of them as critical, though, as the first.

\subsection{The Log4Shell Attack: How it Works}
\label{sub:attack}

The Log4Shell exploit builds upon a JNDI injection vulnerability that was presented at BlackHat in 2016~\cite{mm-jjmrc-16}.
JNDI enables queries of lookup services such as LDAP, the RMI Registry, or the DNS and can load Java objects returned by a service at runtime. Since the query argument is a URL, the lookup can be performed on local or remote services. Via this functionality, an attacker who controls the query can thus load arbitrary code from a location under his control---and this is where Log4j comes in.

Log4j comprises capabilities to interpret strings for enriching logged messages with additional information. Examples are lookups of the Java version or the hostname. These interpreted strings are escaped by wrapping them: \texttt{\$\{prefix:query\}}. In addition to harmless operations, Log4j accepts a prefix that triggers JNDI to perform the lookup, in which case the query includes its own scheme to signify the lookup services used for the query. This extends the known JNDI vulnerability by opening an attack vector via logged messages.
Applications that log web requests, usernames, or generally user-controlled input are easy targets as a result---provided they fail to sanitize their inputs.

\begin{figure}
  \includegraphics[width=\columnwidth]{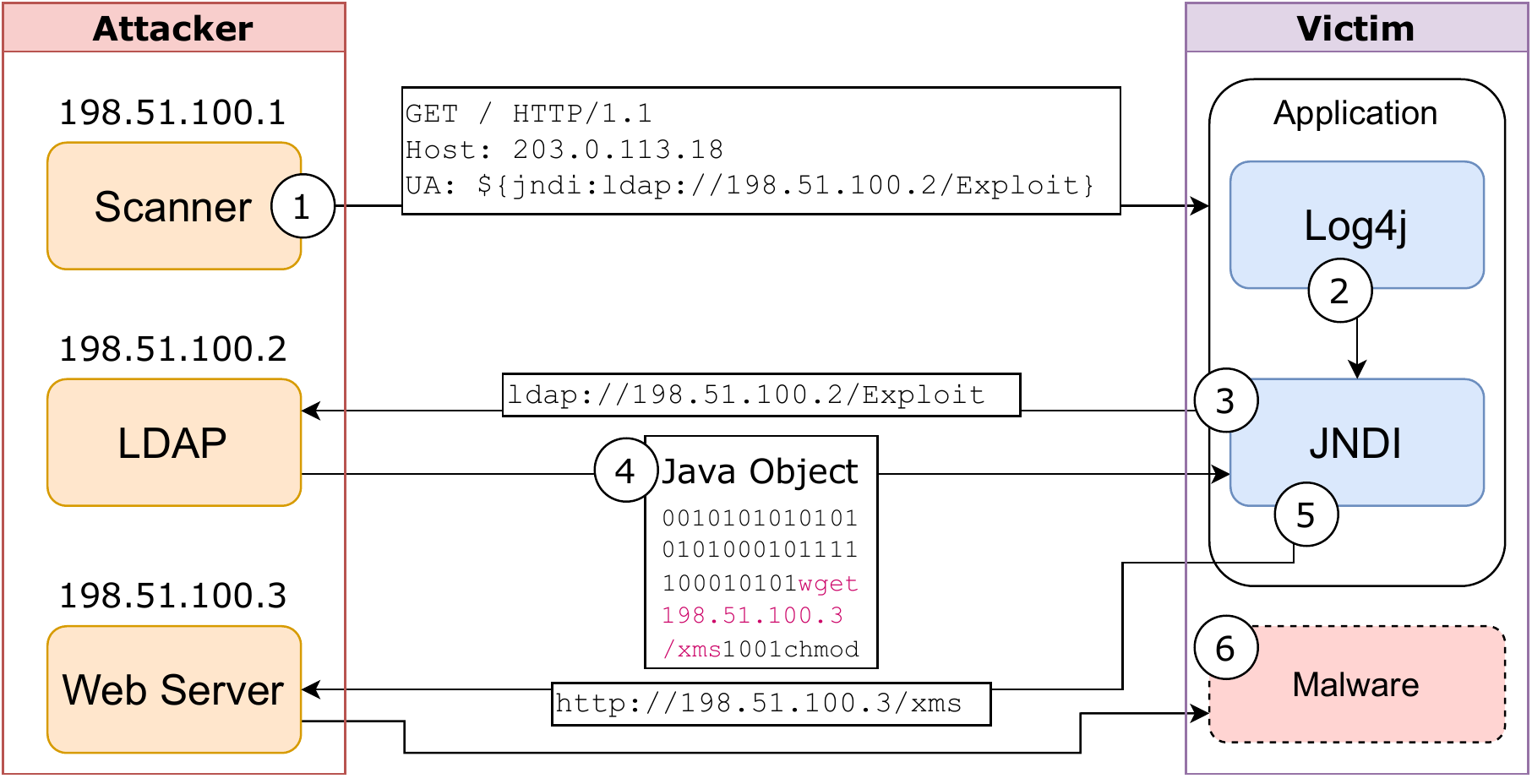}
	\caption{The Log4Shell exploit tricks the application into loading and executing code from a remote server. It is initially instrumented by an executable log string (here ``\texttt{UA:}'').}
  \label{fig:exploit}
\end{figure}

Figure~\ref{fig:exploit} shows a possible exploit scenario initiated by an attacker. The attacker starts with a scan that sends HTTP requests to web servers \circled{1}. Here, the exploit string is placed in the user agent field. An information that operators might log to understand what browsers and operating systems they should support to provide good user experience. In our example, the exploit targets an LDAP lookup via JNDI on an LDAP server controlled by the attacker.

The victim application logs the input string using Log4j~\circled{2}. Log4j in turn finds the escaped string and uses JNDI to perform the LDAP lookup from the remote address \circled{3}. The vulnerable Log4j implementation downloads the Java object prepared by the attacker \circled{4} and loads it locally. This Java object contains a way to run shell commands on the local machine, which downloads the actual malware from a remote server \circled{5} and executes it locally \circled{6}.

Following these steps, an injected payload leads to the execution of arbitrary code prepared by the attacker.
The interpretation of strings by Log4j can further be used to obfuscate the payloads~\cite{m-gpdhe-21}. Instead of including the string ``ldap'' or an address, individual characters are escaped with an operation that leaves them unchanged.

\section{Observing Log4j Scans: Methods and Data Set}
\label{sec:method}

Multiple groups of people hunt for exploitable services: attackers attempting to exploit vulnerable services, researchers who want to examine and analyze, and the security industry, which wants to discover and close vulnerable services before they can be exploited. All of them scan IP space. The time directly after the release of the exploit is the most critical observation period for all, since more and more services are getting patched or taken offline with time.
Scanners that want to identify vulnerable services without exploiting them can simply use LDAP servers which do not perform a valid lookup but only log accessing addresses instead.
This makes it especially hard to attribute malice based on scans alone.

We observe scan attempts targeting TCP on four /24~IPv4~prefixes. Our vantage points neither host services nor are they part of a larger active network. We deploy Spoki~\cite{hnkds-sunws-22}, a reactive telescope that interacts with incoming packets in real-time to establish TCP connections, and collect payloads for a few seconds before closing connections.  Based on the C++ Actor Framework~\cite{chs-ccafs-14} Spoki is highly scalable and 
 can easily handle thousands of addresses and is deployed at all four vantage points. Our vantage points include one in the US (part of the UCSD Network Telescope~\cite{caidaWebsiteTelescope}) and three in the EU. Two of the vantage points in the EU host neighboring networks (EU VP 2 \& VP 3), which are separately announced.
We separate the data set by /24 prefixes in our analysis to accommodate for topological differences. %
We lost data at the US vantage point from Dec~28 to Jan~1.

\begin{figure*}
  \begin{subfigure}{.5\textwidth}
    \centering
    \includegraphics[width=\linewidth]{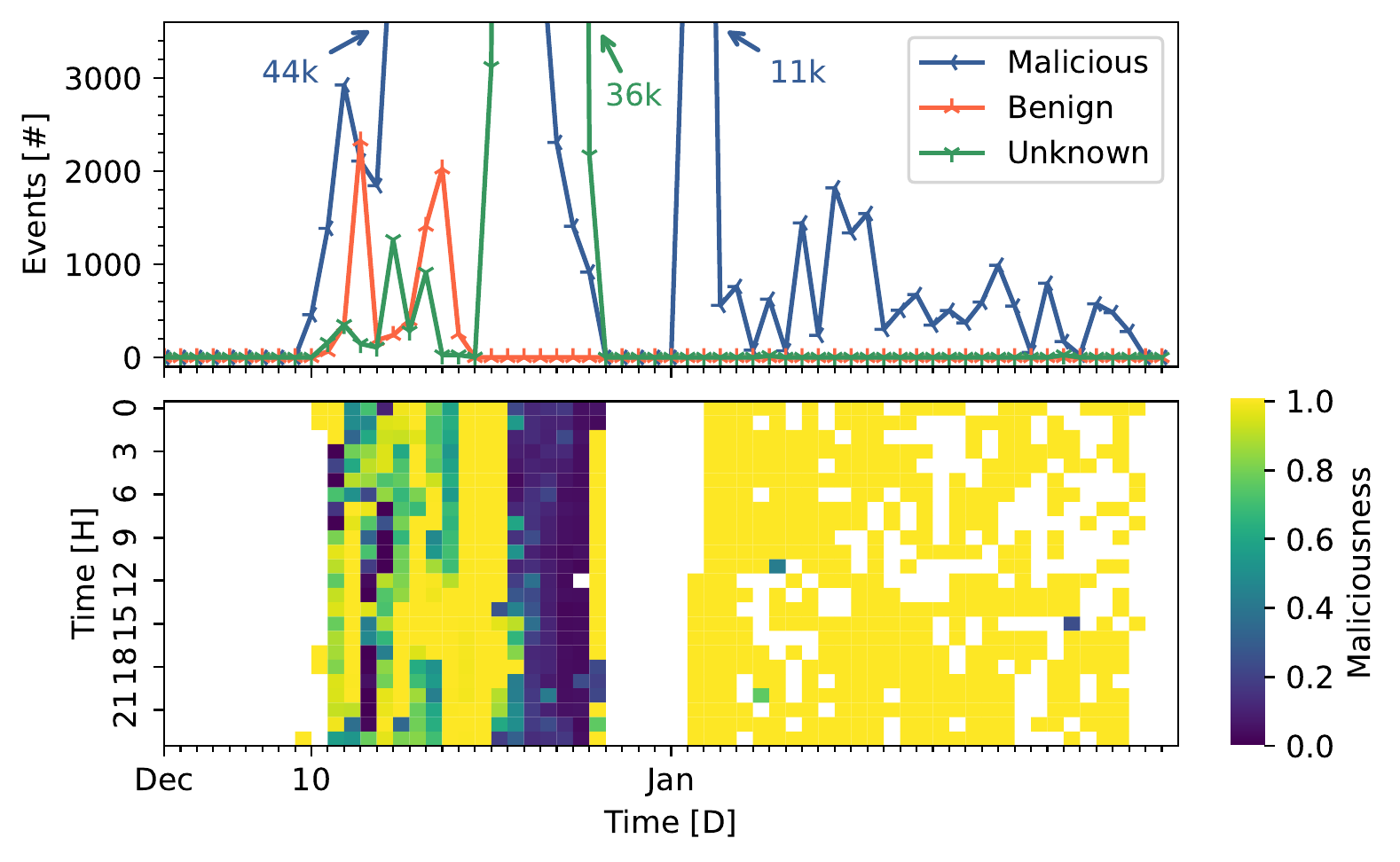}
    \caption{US Vantage Point.}
    \label{fig:maliciousness:ucsd}
  \end{subfigure}\hfill
  \begin{subfigure}{.5\textwidth}
    \centering
    \includegraphics[width=\linewidth]{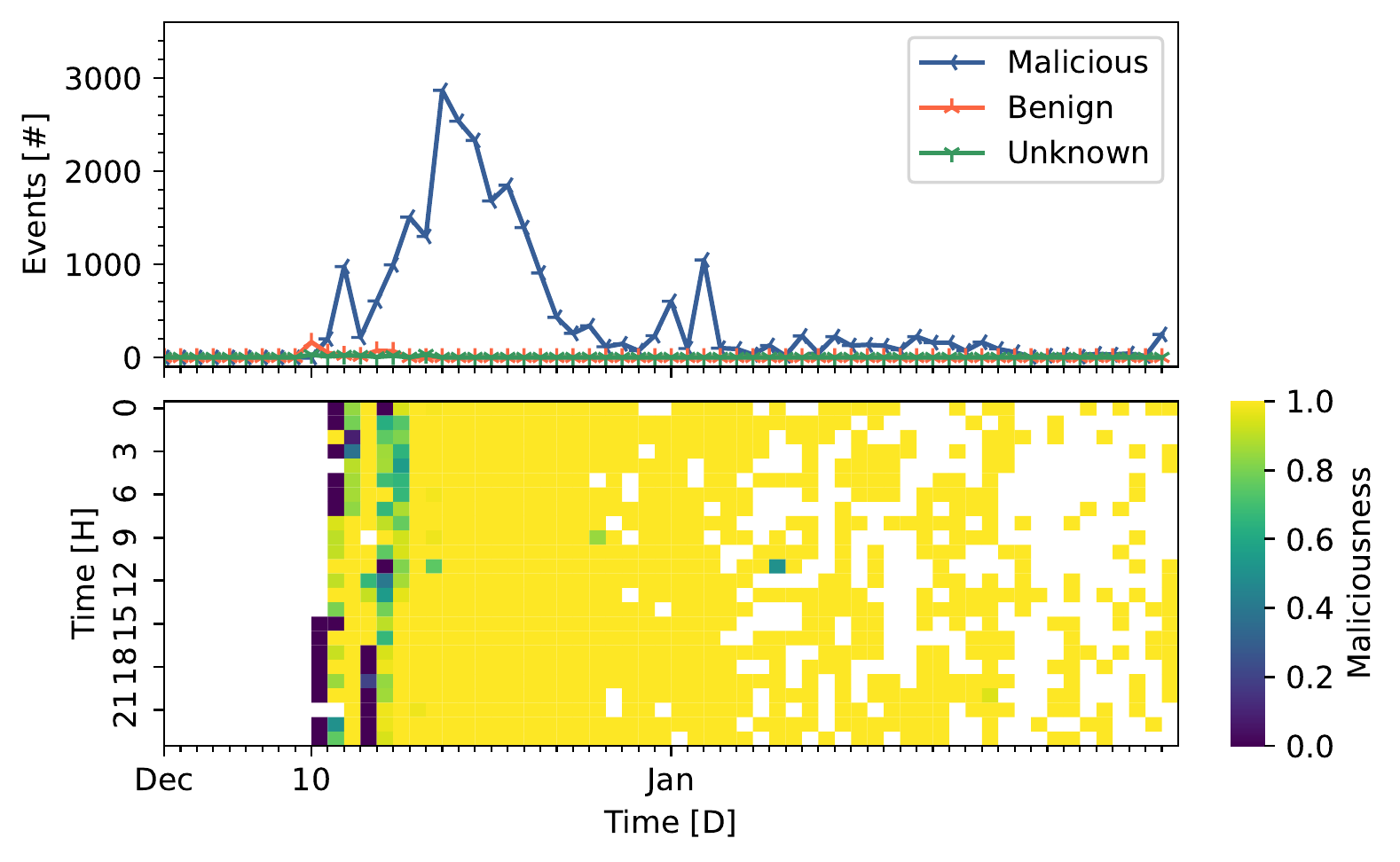}
    \caption{EU Vantage Point 1.}
    \label{fig:maliciousness:bcix:01}
  \end{subfigure}\hfill
  \begin{subfigure}{.5\textwidth}
    \centering
    \includegraphics[width=\linewidth]{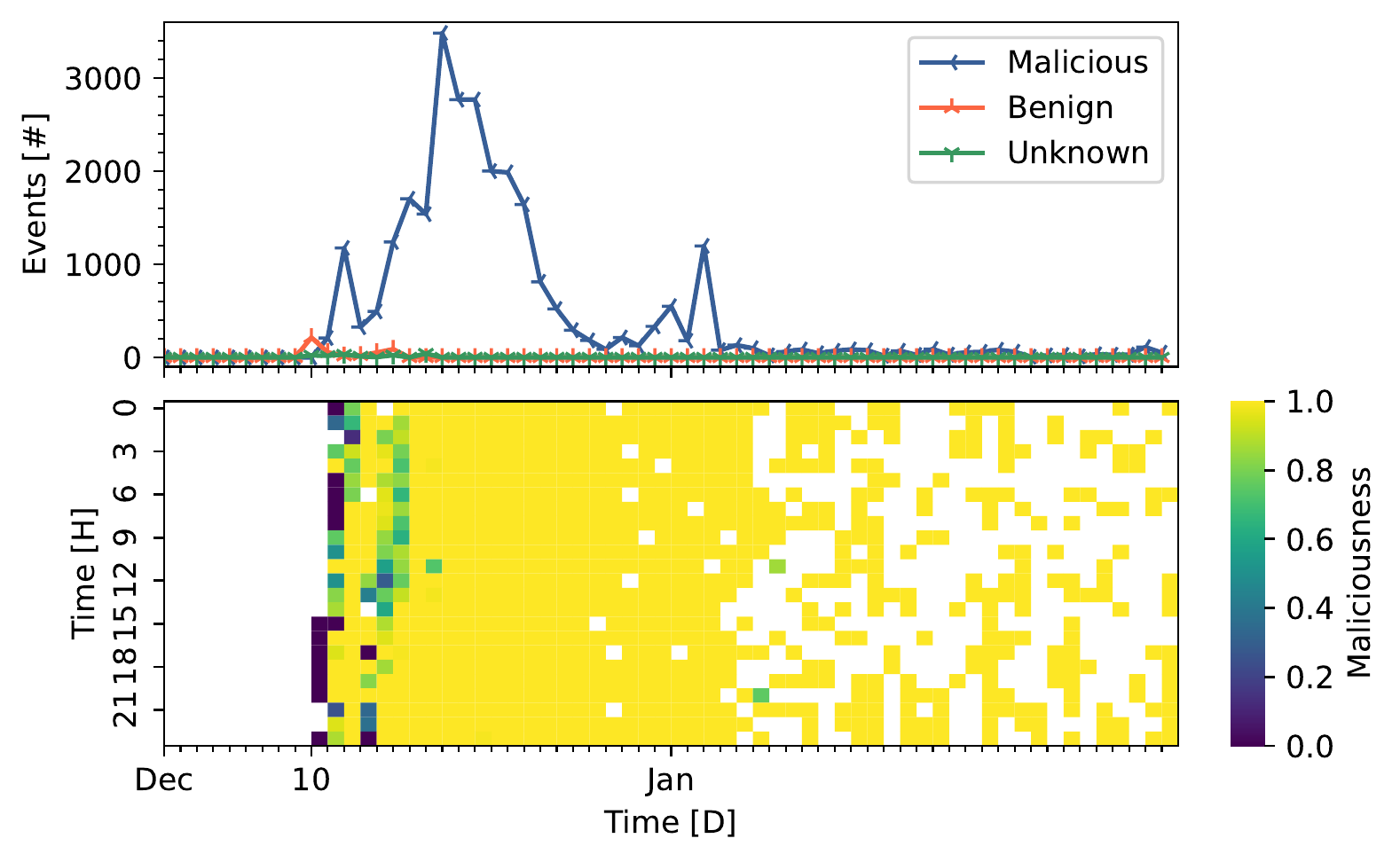}
    \caption{EU Vantage Point 2.}
    \label{fig:maliciousness:bcix:02}
  \end{subfigure}\hfill
  \begin{subfigure}{.5\textwidth}
    \centering
    \includegraphics[width=\linewidth]{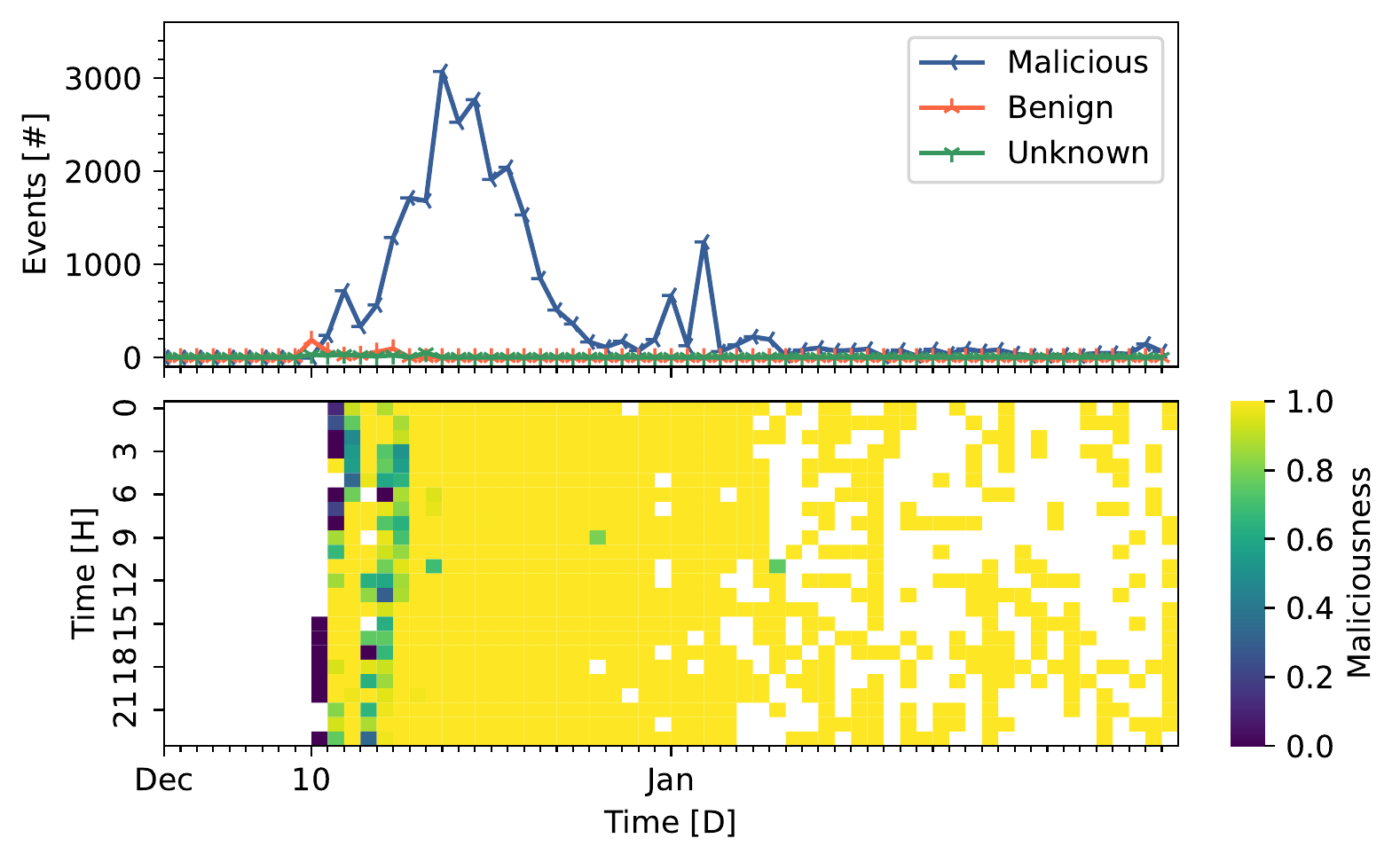}
    \caption{EU Vantage Point 3.}
    \label{fig:maliciousness:bcix:03}
  \end{subfigure}\hfill
  \caption{The intensity and maliciousness of scanners targeting the Log4Shell vulnerability during Dec'21 and Jan'22.}
  \label{fig:maliciousness}
\end{figure*}

Payloads targeting the Log4Shell vulnerability contain the escaped JNDI format string. As such, they are detectable by parsing the payloads if a few obfuscation techniques are taken into account. We use an open source tool~\cite{log4shell-detector} to filter the payloads we collect.
We use MaxMind~\cite{maxmind-geolitecountry} for geolocating IPs. While geolocation is not highly accurate, mapping to countries is still reasonable~\cite{pukdg-igdu-11}. PeeringDB~\cite{peeringdb} gives us an overview over the types of networks we observe and GreyNoise~\cite{greynoise} provides additional threat intelligence information.

\section{Scanners}
\label{sec:scanners}

In this section we examine the scanners of Log4Shell scans. %
These are the sources in step \circled{1}, see Figure~\ref{fig:exploit}.

\subsection{Overview}
\label{sub:maliciousness}

On the day of its disclosure (Dec 10, 2021) we observe the first scans targeting the Log4Shell vulnerability at the EU vantage points. In the US first scans started a day earlier on the eve of Dec 9 (UTC). We collect all events, \ie payloads that include a JNDI exploit string, and classify the sources using GreyNoise~\cite{greynoise} into the categories \textit{malicious}, \textit{benign}, and \textit{unknown}. We cannot easily sort \textit{unknown} sources into the other two categories because Log4j payloads sent during step~\circled{1} do not compromise the system. Still, two payloads that strongly correlate with malicious behavior, see Section~\ref{sec:urls}, were used to further classify  sources as malicious.

Figure~\ref{fig:maliciousness} displays the scan activity over the months Dec'21 and Jan'22. 
The upper graphs show the time series per source type and day while the heat maps below visualize malicious intensity, calculated from the share of malicious events among all events per hour.
The event count in the US graph grows to roughly 44k during the peak of the \textit{malicious} events while the \textit{unknown} events in the subsequent peak first hit roughly 36k.

Mid-December is the period with the highest activity. Noticeably, the only benign events are registered in the first weeks after the disclosure. Towards January the events per day drop to around 100 in the EU, and 2000 or less in the US. The interest in scanning for the vulnerability spiked about a week after the public disclosure. During this time all vantage points receive a strong malicious activity. The US vantage point observes a large share of unknown events as well.

The first scans start on the evening of Dec 9 (UTC+0) at the US vantage point. The EU only sees the first packets nearly a day later (3PM, Dec 10) than the US (11PM, Dec 9). Moreover, the scans in the EU start with scans from BinaryEdge (\url{https://www.binaryedge.io}), a benign threat-intelligence provider---here from DigitalOcean hosts based in the US. At the US vantage point the first scans originate from one source classified as malicious, in a UK AS. %

The heat maps reveal a period of mixed activity before malicious sources take over--with the exception of the unknown sources in the US around mid-December. While the benign actors are fast in the EU, they quickly lose interest. Here, researchers and threat intelligence providers have room to grow: start fast and perform continuous measurements.

From the observed 2023 sources, 1516 were exclusively seen in the US, and 123 only in the three EU telescopes. The remaining 384 addresses were observed at both vantage points. EU VP 1 and 2 bring about $\sim25$ exclusive addresses, EU VP~3 only observed addresses also seen at other telescopes.

\subsection{Event Peaks}

Shortly after the disclosure all vantage points observe benign scanners. At the US VP we observe two spikes up to 2000 packets on Dec 13 \& 18. 90\% of these packets are from a single AS, Alpha Strike (https://www.alphastrike.io). We observe scans focusing on HTTP-related ports such as 8080, 8081, as well as port 8983.

A very high spike in malicious events was observed in the US on Dec 19 \& 20. During these two days, the vantage point records more than 40k packets per day. 80\% of them originate from a single IP endpoint in Russia. This is one order of magnitude more than observed from any other source during this time. All these events target port 8080. %
Two days later, a second spike occurs, this time from unknown sources. During these 5 days the US VP observed more than 30k packets per day. As a result the maliciousness decreases, which is visually striking in the heat map as a dark stripe from Dec 22 to 26. 

Once again, we observe a large share of events (76\%) originating from a single, unclassified IP address of the same Russian hoster of the previous, malicious address that caused the spike days earlier. In contrast to the previous scan these events are split between TCP ports 8080 (18\%) and 5480 (75\%).
A closer inspection of these two events reveals that the first source scans more aggressively but over a shorter time period. %
The hourly packet rate is up to twice as high. In contrast, the unknown source sends at a lower rate over a longer time period, adding up to more than twice as many packets in total. Although their payloads have similarities--they exfiltrate the domain, computer name, OS, and java version--they do not show direct intent to perform RCE via Java objects. We also find no specific overlap in the payload strings and do not see sufficient evidence to classify the source as malicious.%

A malicious spike in events can be observed at all three EU VPs on Dec 18. These spikes reach about 3k events per day. When grouping by AS, two ASes take more than 10\% share: an American hoster and Chinese ISP.
Both scan a range of HTTP-related ports %
as well as ports 5480 and 7547.
These ASes show up at the US VP with a slightly higher event count. Their share is comparatively small due to the high traffic volume from the aforementioned two Russian addresses.

Smaller event spikes from malicious sources are seen at the European VPs on Jan 1 \& 3. These originate from a single hoster IP address. These probes focus on HTTP-related ports. The much larger spike in the US is caused by the same address.

The long tail of smaller spikes in the US originates from a few different ASes. Five make up more than 10\% of the events each, four of which are hosters. Most scans focus again on HTTP-related ports. Among the lesser scanned ports one stands out: 25565, used for MySQL and Minecraft servers.

\subsection{Who is Scanning?}
\label{sub:origins}

\begin{figure}
  \centering
  \includegraphics[width=\linewidth]{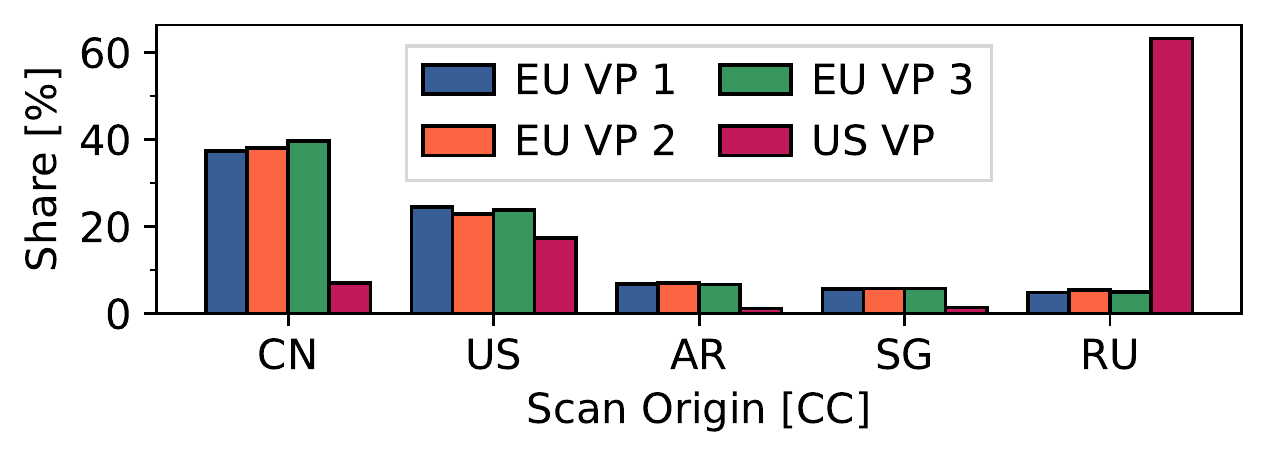}
  \caption{The top 5 countries with at least 5\% share. Russian scanners focus on the US.}
  \label{fig:scanner:geo}
\end{figure}

Geolocation (see Section~\ref{sec:method}) based on the source addresses shows that scans originate to a large extend from three countries. China and the US make up more than 60\% of the events at each vantage point in the EU. In line with the traffic spikes caused by Russian sources in the US, Russia originates more than 60\% of the events observed in the US and around 5\% at the EU vantage points. Figure~\ref{fig:scanner:geo} summarizes the shares of the top five contributing countries.

\begin{table}[b]
  \caption{Overview over the network types from PeeringDB. Hosting providers originate the largest identifiable share.}
  \label{tab:scanner:network:type}
  \centering
  \begin{tabular}{l r r r r}
    \toprule
    & \multicolumn{3}{c}{EU} & \multicolumn{1}{c}{US} \\
    \cmidrule(lr){2-4}
    \cmidrule(lr){5-5}
    & \multicolumn{1}{c}{VP 1} & \multicolumn{1}{c}{VP2} & \multicolumn{1}{c}{VP3} & \multicolumn{1}{c}{VP 1}\\
    \midrule
    Hosting              &   34\% &   34\% &   35\% &   79\% \\
    Transit/Access       &   22\% &   23\% &   22\% &    4\% \\
    Education            &    7\% &    7\% &    7\% &    1\% \\
    Enterprise           & $<$1\% &        &        & $<$1\% \\
    Other                & $<$1\% & $<$1\% & $<$1\% & $<$1\% \\
    Unknown              &   36\% &   35\% &   36\% &   15\% \\
    \bottomrule
  \end{tabular}
\end{table}

To determine the type of network, from which scans originate, we query PeeringDB~\cite{peeringdb}. Table~\ref{tab:scanner:network:type} summarizes the results. 15\% to 40\% of networks do not have an entry or are labeled as \textit{Not Disclosed}.
At all vantage points the largest share falls into the category \textit{Content}, which includes hosting services. Three hosting ASes stand out: Two from the US %
and one from Russia, which predominantly scans the US VP. %

The second-largest share goes to transit and access networks. Traditionally, these are more the home of botnets, which scan from infected hosts~\cite{dkcpp-assb-15,aabbb-umb-17}. Education and business networks tend to scan because of infected host or with the goal to learn  about the new vulnerability.

Educational networks perform scans for research purposes or host compromised machines. The small share of education networks in the US is likely a lack of data in PeeringDB. Although it could further hint that US institutions prefer VMs in the cloud for measurements and infrastructure.

\begin{figure*}
  \begin{subfigure}{.5\textwidth}
    \centering
    \includegraphics[width=\linewidth]{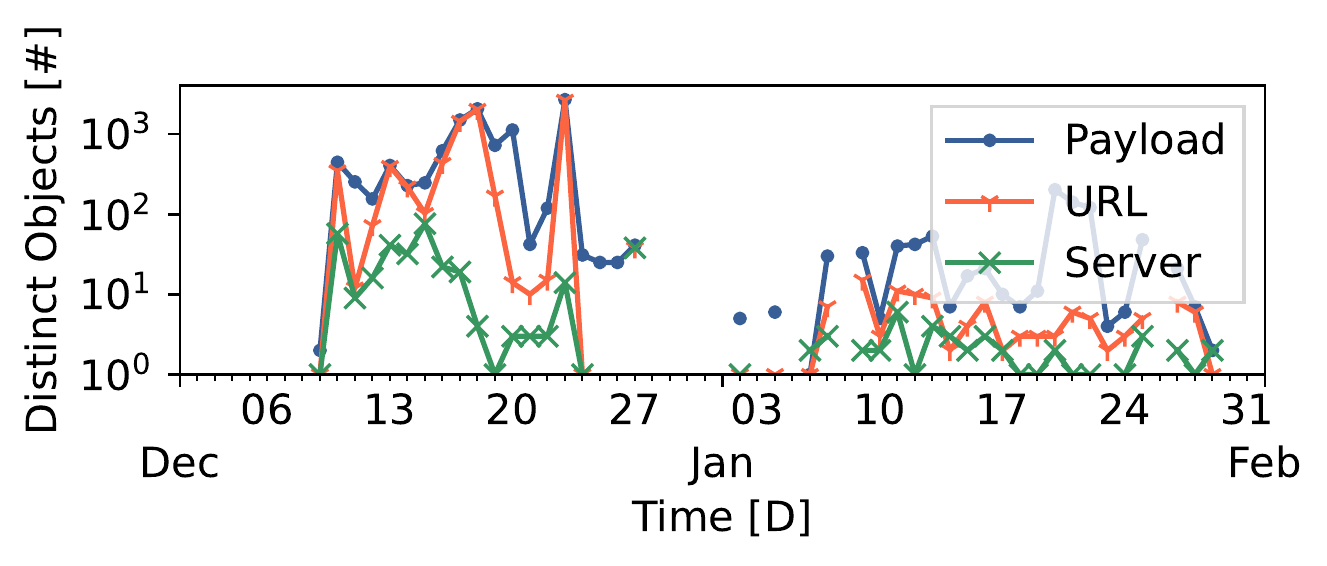}
    \caption{US VP.}
    \label{fig:churn:line:ucsd}
  \end{subfigure}\hfill
  \begin{subfigure}{.5\textwidth}
    \centering
    \includegraphics[width=\linewidth]{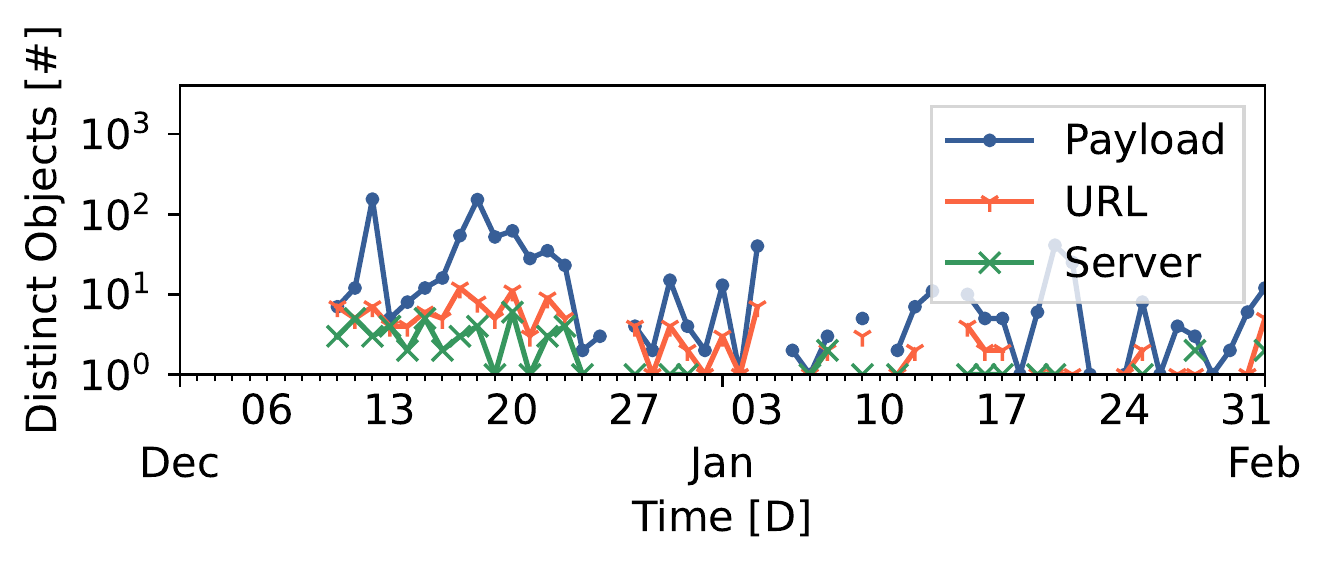}
    \caption{EU VP 2.}
    \label{fig:churn:line:bcix:02}
  \end{subfigure}\hfill
  \begin{subfigure}{.5\textwidth}
    \centering
    \includegraphics[width=\linewidth]{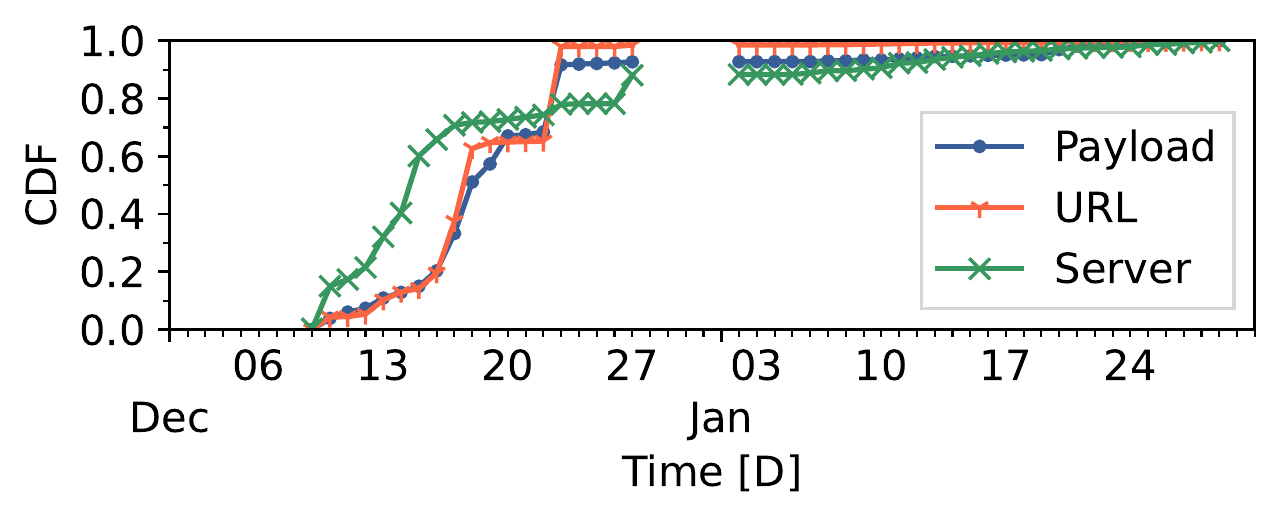}
    \caption{US VP.}
    \label{fig:churn:cdf:ucsd}
  \end{subfigure}\hfill
  \begin{subfigure}{.5\textwidth}
    \centering
    \includegraphics[width=\linewidth]{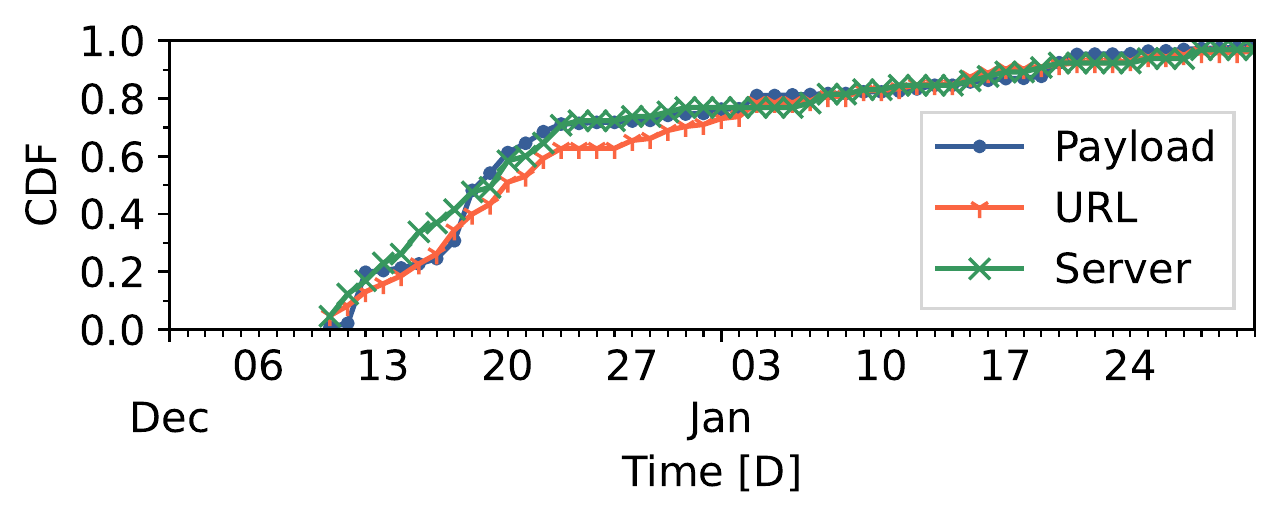}
    \caption{EU VP 2.}
    \label{fig:churn:cdf:bcix:02}
  \end{subfigure}\hfill
  \caption{The relationship between payloads, URLs, and servers over time. (a) and (b) count the distinct objects per day. (c) and (d) aggregate distinct objects over the two-month period in a CDF. EU VPs are represented by EU VP 2 (Dec'21 to Jan'22).}
  \label{fig:churn}
\end{figure*}

\subsection{What are Scanners Targeting?}

The majority of scans aim for HTTP-related ports, such as 80, 8080, 8000. One of the highly-active Russian scanners also focuses on 5480. The three most popular ports account for more than 50\%, likewise the top ten ports account for more than 85\% of the events at each VP. Overall, we observe between 36 and 48 different ports as targets at each VP.

A port that stands out among the top ten in the EU is 7547. This port is associated with the TR-069 vulnerability related to home routers. While this port is frequently scanned (about 4\%), Java is generally not the best fit for home routers and thus Log4j is not likely to be used in such an environment. Presumably, scanners are just testing the port in passing.

Scan packets do not uniformly distribute across the addresses of our prefixes. At every vantage point the difference between the least scanned address and the most scanned address is at least a factor of 1.5 (US) and 2.5 (EU).

\section{Payloads of the Scanners}
\label{sec:payloads}

All observed Log4Shell events contain a payload within the  JNDI lookup string. Hence, the number of payloads we collected equals the number of events described in Section~\ref{sec:scanners}. The basic idea of such payloads with embedded exploits was explained in Section~\ref{sub:attack}. We now analyze the collected payloads, \ie the data scanners send in step \circled{1}, in detail to learn about what they impact, how payloads change over time, which protocols scanners use, and the JNDI URL placement.

\subsection{Temporal Development}

Section~\ref{sub:maliciousness} revealed how scans reached a quick peak before they rapidly decline and settle in a low volume. We cannot tell from the event counts alone whether scanners keep their setups and continue to run unaltered, \ie reusing payloads, or whether they purposefully rotate payloads. The variation in payloads over time can help to answer this question.

Figure~\ref{fig:churn} depicts the evolution of distinct payloads, URLs, and hosting servers over our observation period. Subfigures~\ref{fig:churn:line:ucsd} and \ref{fig:churn:line:bcix:02} show the distinct counts per day. In contrast, Subfigures~\ref{fig:churn:cdf:ucsd} and \ref{fig:churn:cdf:bcix:02} show the CDF of distinct objects over the complete period. The payloads are the data we receive directly via TCP, URLs are the escaped JNDI URLs in the payloads, and servers are solely the addresses in those URLs, \ie the host and port information of LDAP or RMI server controlled by the attacker. Because payloads and URLs often include the address of the victim, \eg in the host parameter of the HTTP header, we replace the IP addresses from our subnets with a static string. This way, payloads that only differ by their destination will coincide.

The event spikes in Figures~\ref{fig:churn:line:ucsd} and \ref{fig:churn:line:bcix:02} correlate with the spikes about a week after the disclosure, see Figure~\ref{fig:maliciousness}. Note that the graph has a logarithmic y-axis. At both vantage points the distinct payloads count is much higher than the server count. Scanners use different payloads for the same injection string---likely to test the behavior of the scanned service as it might differ by application which data is logged. Similarly, the number of distinct URLs is higher than the number of distinct server addresses. In those cases, attackers re-use the same server, but with different paths. In practice this means that individual servers are used to distribute multiple types of downloads or malware. On some days the number of payloads equals the URL count. Here, scanners encoded additional data in the URL, such as a HEX-encoded JSON payload that includes the destination and other information.

We now analyze the evolution of the attack ecosystem over the complete time frame, \ie the new payloads, URLs, and servers.
Figures~\ref{fig:churn:cdf:ucsd} and \ref{fig:churn:cdf:bcix:02} plot the corresponding CDFs. Noticeably, the behavior of distinct payloads and infrastructure largely differ between the US and the EU vantage points: 
The US graph payloads and URLs has more extreme steps, one from Dec 17 to 20 and one from Dec 22 to Dec 23. These jumps lead up to the high spikes in malicious and later in unknown events. In contrast, the server line rises much earlier. The rise in payloads and URLs does coincide. This indicates varying attacks in the US backed by the same infrastructure---supposedly, attackers adapted their payloads and URLs to explore the attack surface.

At the EU VP, all three measures occur in line. There are two steeper increases in payloads shortly after mass scanning started on the Dec 12 and again on Dec 18. Still, we do not see a divergence as in the US. Payloads, URLs, and servers behave similarly, rising on similar days.
Scanners vary their payloads. Most frequently we observe classic horizontal scans and scanners sending different payloads to a single endpoint before moving on.

\begin{figure*}
  \begin{subfigure}{.5\textwidth}
    \centering
    \includegraphics[width=\linewidth]{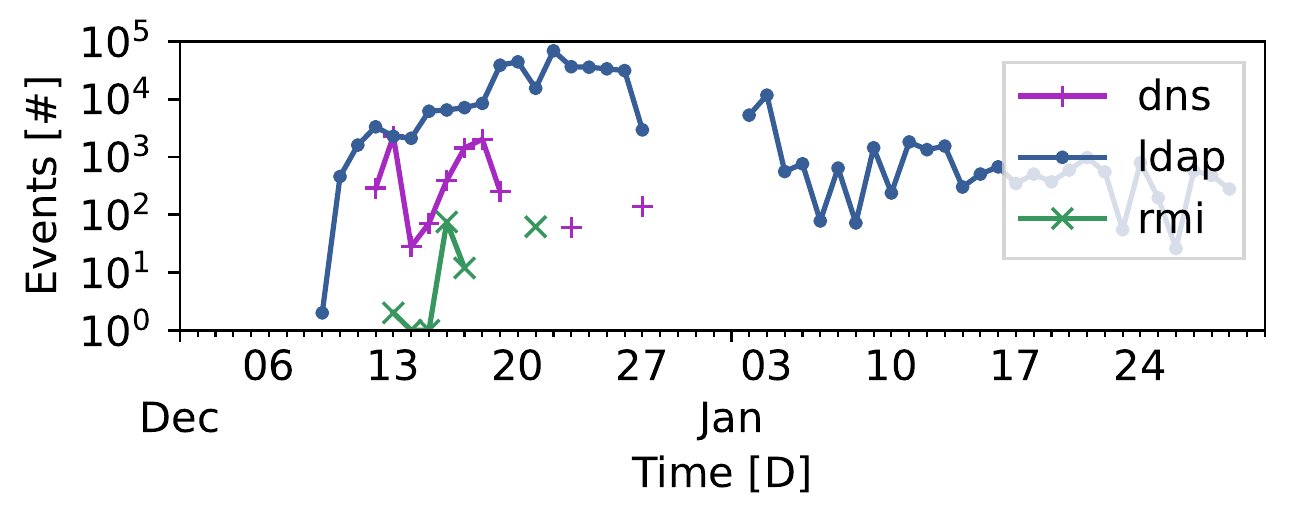}
    \caption{US VP.}
    \label{fig:schema:ucsd}
  \end{subfigure}\hfill
  \begin{subfigure}{.5\textwidth}
    \centering
    \includegraphics[width=\linewidth]{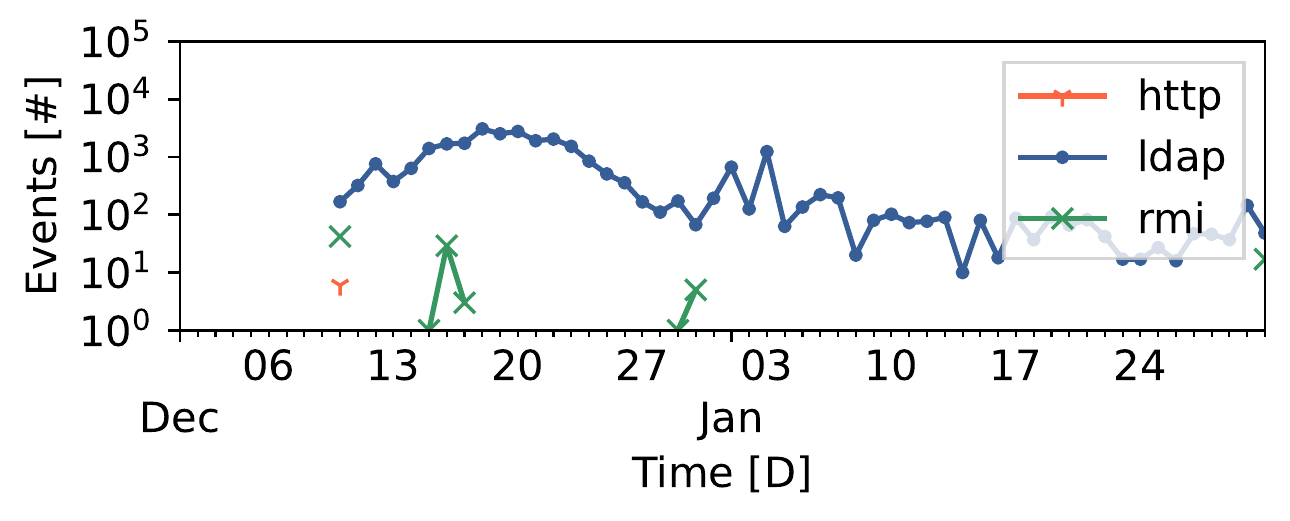}
    \caption{EU VP 3.}
    \label{fig:schema:bcix:03}
  \end{subfigure}\hfill
  \caption{Schemes in JNDI URLs over time. LDAP is used nearly exclusively (Dec'21 to Jan'22).}
  \label{fig:schema}
\end{figure*}

\subsection{Exploit String Placement}

While focusing on HTTP related ports we observe that payloads of the scanners are either GET (between 91\% and 98\%) or PUT requests (remaining). Note that we cannot observe all application-specific payloads as they may be encoded in an application-specific format. Nevertheless, binary protocols that contain a matching string in ASCII would still be detected.

Exploit strings in payloads become visible at various locations. This reflects attackers who still search for the best attack method, as well as attackers that try to alter their payloads continuously to evade detection. The most popular placements in HTTP headers are summarized in Table~\ref{tab:url:location}.

Regularly logged locations such as the \texttt{User-Agent} (UA) and \texttt{Authentication} (Auth) header fields are popular targets. The UA, which can be customized for specific purposes~\cite{g-gcua-22}, identifies the client. Websites depend on it to deliver matching content. As a result, it is often logged to keep statistics on users. Authentication information are similarly important for debugging and access control. Since HTTP header manipulation is a known issue~\cite{thccp-ehhmi-17}, logging any header fields should be done with care.
 
\setlength{\tabcolsep}{3pt}
\renewcommand{\arraystretch}{0.8}
\begin{table}[t]
  \caption{The most common header locations to store the JNDI URL. (UA: \textit{User-Agent}, Auth: \textit{Authorization}, X-Api-Ver: \textit{X-Api-Version})}
  \label{tab:url:location}
  \centering
  \begin{tabular}{l@{\ \ } l@{\ } r l@{\ } r l@{\ } r l@{\ \ } r}
    \toprule
    & \multicolumn{6}{c}{EU} & \multicolumn{2}{c}{US} \\
    \cmidrule(lr){2-7}
    \cmidrule(lr){8-9}
    & \multicolumn{2}{c}{VP 1} & \multicolumn{2}{c}{VP2} & \multicolumn{2}{c}{VP3} & \multicolumn{2}{c}{VP 1} \\
    \midrule
    1. & UA    & 23\% & UA    & 22\%
       & Auth. & 23\% & UA    & 11\%
    \\
    2. & Auth. & 20\% & Auth. & 21\%
       & UA    & 22\% & Path  &  9\%
    \\
    3. & Path  & 13\% & Path   & 14\%
       & Path  & 14\% & Cookie &  6\%
    \\
    4. & Cookie & 10\% & Cookie & 11\%
       & Cookie & 11\% & Auth.  &  6\%
    \\
    5. & X-Api-Ver.    & 10\% & X-Api-Ver.    &  8\%
       & X-Api-Ver.    &  9\% & Referer       &  6\%
    \\
    \bottomrule
  \end{tabular}
\end{table}

The lower shares in the US stem from the high variety in header configurations and fields sent by the Russian scanners. While we saw around 45 different header fields in the EU, the US VP observed 157 distinct fields. During January the field count shrinks to 13, favoring User Agent and X-Api-Version.

The two Russian scanners follow different scanning styles. The one IP address tagged as malicious appears to partly randomize the headers. All payloads are GET requests that include the user agent \texttt{Go-http-client}, a \texttt{Host} field, the ``randomized'' field with the JNDI string, and end with \texttt{Connection: close}. Exception are payloads that contain the JNDI string as the user agent, in which case only the \texttt{Host} field follows. Not only do they change their fields but choose from several simple obfuscated URLs.

In contrast, the other Russian scanner rotates payloads systematically. It sends the same payload to a number of addresses before changing a single characteristic, such as the header field that contains the JNDI string or the obfuscation method. It looks like a nested loop that rotates the obfuscation in the inner loop and the header field in the outer one. The user agent is only included when it contains the JNDI string.

The obfuscation approaches we observe are built on the substitution used for the vulnerability itself. There are a few easy obfuscations that hide the strings ``jndi'', ``ldap'' or other parts of the URL, hiding keywords that can otherwise be found by simple pattern matching, thus making the payloads harder to detect. As an example, the string \texttt{\$\{lower:j\}} will substitute to ``j''. As the obfuscation is very easy to achieve and the general JNDI exploit is not new, obfuscated payloads quickly appeared in our data. At the EU vantage points, the first obfuscated payloads showed up shortly after midnight on Dec 11. In the US it took a day longer.

\section{Examining the JNDI/LDAP Exploitation}
\label{sec:urls}

In this section, we investigate the final steps \circled{3} to \circled{5} of the attack, namely the malware requests triggered via JNDI. We also acquire and inspect the malware distributed via Log4Shell. The URLs used to query via JNDI have four parts: a scheme, a host, a port, and a path.

\subsection{Analyzing the URLs}
\label{sub:fragments}

\paragraph{Schemes}
Tricking the victim into downloading Java objects via JNDI is central to the exploit, see step \circled{3}. JNDI supports a range of services, but only a few are used for Log4Shell. 

Figure~\ref{fig:schema} shows the event count for each scheme we observed (note the logarithmic y-axis). The EU vantage points see LDAP almost exclusively. RMI occurs a handful of times at both vantage points. Aside from the events on Jan 16, which originate from the same ASes, there is no correlation between the VPs. As a third scheme, EU VP 3 observed a single HTTP request. The website behind the address gives security advice on how to patch Log4j (Feb'22). The name in the reverse DNS record of the scanning node matches BinaryEdge. In the US we observe two spikes of DNS schemes in January. 95\% of these originate from the Alphas Strike AS, matching the benign spikes in Figure~\ref{fig:timeline}.

Aside from LDAP, attacks over RMI received attention in the media~\cite{j-lvasf-21}. We only observed three (US) request with the URL mentioned in the article. These attacks may have been focused and did not hit our VPs in bulk.

\begin{figure}[b]
  \centering
  \includegraphics[width=\linewidth]{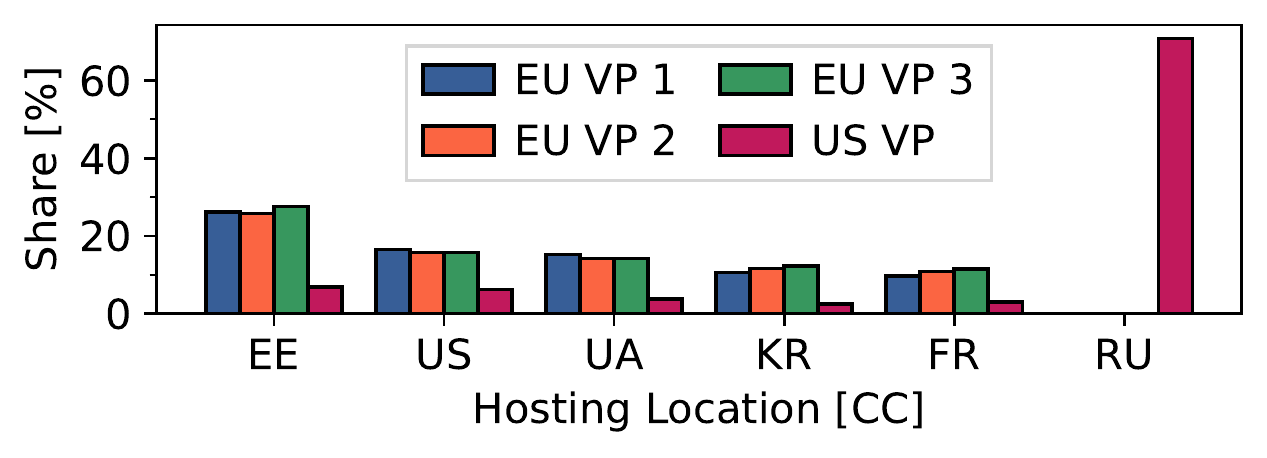}
  \caption{The top 5 countries that host Log4Shell servers. Russia only takes a share in the US.}
  \label{fig:hoster:geo}
\end{figure}

\paragraph{Hosts}
PeeringDB lacks information about nearly half of the hosting networks we observe. We notice that corresponding servers are mostly located in two ASes, both from Estonia (one provider shows up with two ASes, one of which is Ukrainian). Manual search identified both as hosters. Relabeling them accordingly, the servers we observe are mostly located in what PeeringDB labels as \textit{Content} ASes, \ie hosters (EU: 70\%, US: 80\%). The large share of hosters fits the distribution model for malware. These servers need to be reachable from everywhere with a high uptime to allow compromised machines to download malware. Transit and access networks together make up roughly 20\% at the EU VPs and 5\% at the US VP. 

The most popular location for servers are displayed in Figure~\ref{fig:hoster:geo}. While Estonia runs top in the EU, Russia takes the biggest chunk for the US by far. Notably, none of the Russian servers were observed in our EU vantage points.

Next we compare the location of the scanners to the server locations. In the EU region the most common combination is scanning from China and hosting in Estonia (about 15\%). Second rank is using the US for both scanning and hosting (roughly 12\% of the events) These combinations are dwarfed by the Russian scanners observed in the US (70\%), which use the same IP for scanning and hosting the LDAP server. Here, the combinations US and US (5\%), China and Estonia (3\%) follow on places two and three.

\paragraph{Ports}
Less than two percent of the LDAP servers used for the attacks are bound to the default LDAP port (389). Instead, the port most commonly used is 1389. The share ranges from 93\% to 96\% at all vantage points. %
Other ports in use are 2420 and 80 in the EU, each between 1\% and 2\%. In the US we observe 12344 at around 2\% as well. Such a high share could hint at common tools or tutorials used by the attackers.

\paragraph{Paths}
Except for a single path, observed paths do not conform to the RFC that defines them~\cite{RFC-4516} as they do not include a valid distinguished name. %
Two paths stand out among the LDAP URLs. First is \texttt{/Exploit}, which makes up the largest share at all vantage points with 70\% to 80\% in the EU and 20\% in the US. A second group of paths shares the segments \texttt{Command/Base64} followed by a base64-encoded segment. This group takes the second-largest share. These paths begin in a variety of ways, potentially hinting at their purpose, such as \texttt{TomcatBypass} or \texttt{GroovyBypass}. Decoding the base64 segment reveals script code (mostly bash) that downloads an executable via HTTP to run locally.

At first glance, encoding commands in a URL path is an odd choice. However, the second piece of the puzzle is an LDAP server implementation, which dynamically builds Java objects that will run the base64-encoded command. These servers can be found on GitHub in several repositories, although the original\footnote{\texttt{https://github.com/feihong-cs/JNDIExploit}} is no longer available. This server---aptly named \texttt{JNDIExploit}---binds local ports for LDAP and HTTP. It responds to a variety of queries that include the \texttt{Base64} fragment and match the paths we observed. The most common variant we see is the \texttt{TomcatBypass}.

Having such tooling at hand makes it easy for attackers to set up the attack and run it. We confirmed that servers used in attacks exhibit this dynamic behavior by sending a custom base64-encoded command. We received the Java object in response of our custom code.
The \texttt{JNDIExploit} repositories use port 1389 as the default for LDAP, which matches our observation of common LDAP ports in the URLs.

\subsection{Downloading Malware}
\label{sub:malwaredownload}

The Java objects are just an intermediate step to the goal of the attack: infecting the host with malware. We follow this path to find out which malware is distributed this way. %
When running the downloader in Feb'22, most servers were no longer available. We successfully download nine distinct Java objects, compare step \circled{4}.

The LDAP answers contain two important keys: \texttt{javaClassName} and \texttt{javaSerializedData}. In all cases the class name is set to \texttt{java.lang.String}. The objects we collected match the objects built by the JNDIExploit LDAP server, see Section~\ref{sub:fragments}. One was built for the ``Groovy bypass'' and the remaining were built for the ``Tomcat bypass''. The serialized objects look like a Java \texttt{StringRefAddr} object. 

Both bypasses use slightly different ways to run code. The Tomcat bypass instantiates a script engine to run JavaScript code while the Groovy bypass builds on Groovy itself. The script code is encoded in the serialized objects as ASCII. One of the Tomcat samples executes PowerShell code, which likely targets Windows. While most of the other scripts include a mechanism to determine the local OS, e.g., by checking the direction of slashes in a file path, they execute bash commands either way, and are thus unlikely to run on Windows.

We extract the download commands in our small samples set by hand and download three different binaries, compare step \circled{5}. (Four connections failed and three of the objects contain the same scripts, although they differ slightly.) The hashes of all downloads are registered on VirusTotal~\cite{virustotal}, where they were first submitted between the mid and end of Jan'22, i.e., while Log4Shell attacks were taking place. Two of the samples are scripts while one is an 32-bit ELF binary.

The malware acquired via the PowerShell code is itself a PowerShell script and downloads a binary with the name of a known crypto miner. Although the other shell script looks more sophisticated---it stops local programs, downloads an uninstaller to remove software, adds new cron tab entries, and tries to make its way into connected hosts via ssh---its goal is similar: installing a crypto miner.

Understanding the base64-encoded commands opens another way to acquire malware. Instead of taking the round-trip via the LDAP server, which embeds them into a Java object, the commands can be decoded directly. Via this method we can acquire an additional binary: a 64-bit executable ELF file first registered on VirusTotal at the beginning of Dec'21. The low yield matches the churn in URLs we observe in Section~\ref{sec:payloads}.

\subsection{Locating Malicious LDAP Servers}

We finally explore the malicious server infrastructures by active scans. To this end, we utilize the fact that the Log4Shell exploit requires publicly reachable servers that return Java objects, compare step \circled{3} and \circled{4} in Figure~\ref{fig:exploit}. %

Malicious servers predominantly listen on port 1389 coupled with a non-standard LDAP behavior. This is why we use ZMap~\cite{dwh-zfiws-13} to scan TCP/1389 for open ports. We then identify unsecured LDAP servers by performing LDAP bind operations, which should fail on servers enforcing authentication. In a next step we query for the two most common paths observed: \texttt{/Exploit} and a path with \texttt{Base64} string. 

We find 5.1M servers responding to SYNs, but only 1,110 allow unauthorized LDAP-binding. 81 servers return answers for \texttt{/Exploit}, and 179 for \texttt{Base64}. These sets overlap, which leads to 183 malicious LDAP servers in total (16\%). Comparing to Figure~\Ref{fig:churn}, we infer that the number of daily used and dormant servers differ by an order of magnitude.

We collect six Java objects via the \texttt{/Exploit} path and 97 objects via the base64 path. Their general structure matches the objects from Section~\ref{sub:malwaredownload}. Since base64 requests are expected to encode our command they should not return any malicious objects. Six objects are an exception. Here, servers return the same payload for both paths, likely a static response. We identify one downloader for a 64 Bit ELF binary, an object that runs PowerShell code, and one broken object. The remaining three payloads do not include script code.

\section{Discussion and Conclusion}
\label{sec:conclusion}

The ease with which Log4Shell can turn into exploits is part of what make it as critical as it is.
Access to some public interface suffices to abuse it. YouTube videos explain details and give guidance on how to apply the tools. %
The underlying problem is input sanitization---a common challenge that has plagued the industry in the form of SQL injections for years. User-supplied input must be treated with caution.

On a positive note, the vulnerability saw wide coverage online as blog posts, lists of vulnerable applications, and detection tools were quickly published.
At the same time official organizations published reports and issued warnings.

In this paper, we observed Log4Shell scanning through reactive network telescopes immediately after the disclosure. Most noticeable were the large spikes of malicious events about a week after. These hit all vantage points but particularly targeted the US, giving the scans a geographical focus. More vantage points are necessary to confirm this as a global pattern.

Our analysis showed common characteristics such as two prevalent path fragments: \texttt{Exploit} and base64-encoded commands. Such similarities hint at shared tools or tutorials. This assumption is further strengthened by a common LDAP port we observed during the majority of attacks and active scans, which aids us in identifying LDAP servers used for attacks.

Although the overall scan rate decreased significantly during our observation period, malicious scanners continue to scan for the vulnerability and transport new payloads.

The methodology presented in this paper is limited to the observation of scans. As such, we cannot measure the success rate of attackers, but observing scanning behavior can be an expressive indicator for the liveliness of the scene. 
A quick decrease in scanning activity may also reflect that vulnerable services are thinning out.

While the list of affected software contains many popular applications~\cite{ncsc-log4shell}, the long term effects are yet unclear. Many applications quickly saw patches, but their rollout will eventually slow down. Due to the many attack vectors it is hard to find all active vulnerable applications using a single methodology. The combination of complementary approaches will be part of our future work.

\section*{Acknowledgments}
We thank CAIDA for providing access to the UCSD Network Telescope and the network operators who support us deploying our telescope in Europe.
We appreciate fruitful discussions with Johannes Klick and Stephan Lau.
This work was partly supported by the German Federal Ministry of Education and Research~(BMBF) within the project \emph{PRIMEnet}.

\balance
\bibliographystyle{IEEEtran}
\bibliography{cites, internet, rfcs, own}

\label{lastpage}

\end{document}